# Enhancement of Dual-Band Reflection-Mode Circular Polarizers Using Dual-Layer Rectangular Frequency Selective Surfaces

M. Fartookzadeh and S. H. Mohseni Armaki

*Abstract*— A new kind of dual-band reflection-mode circular polarizers (RMCPs) is introduced with wide bandwidth and wide-view at the operating frequencies. The proposed RMCPs are based on dual-layer rectangular patches on both sides of a substrate, separated by a foam or air layer from the ground plane. Required TE susceptance of the first layer patches to produce circular polarization is calculated using the equivalent transmission line model. Dimensions of the RMCP are obtained using parametrical study for the two frequency bands, 1.9-2.3 GHz and 7.9-8.3 GHz. In addition, it is indicated that the accepted view angle and bandwidth of the proposed dual-layer RMCP are improved compared with the single layer RMCP, significantly. Moreover, a tradeoff is observed for the dual-layer RMCP on the bandwidths of X band and S band that can be controlled by propagation angle of the incident wave. The proposed RMCP has 30.5 % and 33.7 % bandwidths for less than 3 dB axial ratio with incident angles $\theta_{max} = 50°$ and $\theta_{min} = 35°$. Finally, simulation results are met by the measurement for three angles of the incident wave.

## I. INTRODUCTION

Circularly-polarized antennas are widely spread on the radio wave applications. Radar systems, earth stations, scanners and satellite systems are examples of these applications. Circular polarization (CP) can be obtained by the circularly polarized antenna elements using regular methods such as multi-feed antennas, helical antennas, spiral antennas, etc. [1-6]. Another method is the sequential rotation technique that can obtain the CP from the array of linearly polarized antennas with unique angular and phase arrangements [7- 9].

Reflection-mode circular polarizers (RMCPs) and transmission-mode circular polarizers are alternative methods to produce CP from linearly polarized antennas. Various configurations of these polarization converters are introduced so far. Parallel plates, patches and lines, meandered lines, metallic helices, and U-shaped split ring resonators [10-17] are examples of transmission-mode circular polarizers. For RMCPs artificial magnetic conductors [18], arrays of L-patterns [19] and dipole (patch) array structures [20, 21] can be named.

Furthermore, several methods have been suggested for designing dual-band polarization converters, such as the structures based on planar spirals [22], split ring resonators [16, 17, 23], a pair of two-turn spiral resonators [24], etc. RMCPs consisting patch arrays are also applied for dual-band application with two sizes of patches for two desired frequency bands. The small size patches can be placed beside large ones [25] or inside them using slots for separation [26]. However, all of the proposed RMCPs for dual-band applications are affected by fundamental limitations on the bandwidths at their operating frequencies. The RMCP with two size patch arrays [26] has wide bandwidth for the first operating frequency, however the second operating frequency has a narrow bandwidth and low permitted view angle. In this paper, a new approach is used to design dual-band RMCPs for improving the bandwidth of second operating frequency from 9.8 % in [26] up to 33.7 %. In addition the view angles are improved from $\theta_{max} = 39°$ and $\theta_{min} = 25°$ in [26] to $\theta_{max} = 50°$ and $\theta_{min} = 35°$.

The proposed RMCP is consisting of dual layer rectangular patch arrays. Patch arrays are placed on both sides of a substrate, separated by a foam or air layer from the ground plane. Equivalent transmission line model of the structure is used to obtain required TE susceptance of the first layer for CP. Required TE susceptance curves are plotted for different sizes of the patches on both layers. TE susceptance curve of the first layer should pass from the required curves on the desired frequency bands. The desired frequency bands are 1.9-2.3 GHz and 7.9-8.3 GHz [26, 27]. Therefore we can increase the angle difference (beamwidth) from 14° in [26] to more than 20° by using the proposed RMCP. Moreover, the dimensions can be scaled and tuned for other frequencies explicitly.

Subsequently, dimensions of patches are obtained using a parametrical study and observing the effects of changing dimensions on the susceptance curves. We observed some differences between simulated and predicted results of the RMCP that is construed as the coupling effects between layers. However the effects of changing sizes are similar in calculation and simulation. Therefore dimensions of the final structure are obtained by minor changing of two parameters of the initial RMCP. The calculation method and parametrical study are explained in next section. Simulation results for different angles of incident wave are discussed in section 3. In addition, simulations are validated by the measured results for some angles of the incident wave.

## II. DUAL-LAYER RMCP CONFIGURATION

Equivalent transmission line circuit model of the proposed RMCP is indicated in Fig. 1. The relative permittivity of substrate is 3.9 for the first layer and the air substrate is used for the second layer. For the RMCP with single layer frequency selective surface (FSS) [26], $Y_{s2}$ vanishes and therefore we have $Y_{i2} = Y_{in2}$. Dual band operation for this RMCP is obtained using two sizes of patches each for one operating frequency as indicated in Fig. 2 (a). The small patches are placed inside the large patches using slots for separations. In addition, the effects of small patches are neglected in the analytical method about 2 GHz and the effects of large patches are neglected about 8 GHz.

The RMCP with dual-layer FSSs is consisting of only rectangular patches on both layers. The sizes and separations of patches are obtained using the equivalent transmission line circuit and simulation results. Schematic of the proposed

M. Fartookzadeh and S. H. Mohseni Armaki are with Department of Electrical and Electronics Engineering, Malek Ashtar University, P. O. Box 1774-15875, Tehran, Iran.

RMCP unit-cell is indicated in Fig. 2(b) and the array with slashed first layer is indicated in Fig. 2(c). It is assumed that the unit-cell of second layer is five times greater than the first layer in $\hat{x}$ direction and both layers have similar dimensions in $\hat{y}$ direction. In particular it is assumed that $l_1 = 3.5$ mm, $l_2 = 17.5$ mm, $w_1 = w_2 = 3.75$ mm. These dimensions are obtained from initial calculations of the required TE susceptance of the first layer ($B_{s1}^{TE}$) and the fact that the TE susceptance of patch array as function of frequency is a line that passes from the origin [26, 28]. The effect of other parameters will be indicated shortly. In addition it will be observed that the sizes of RMCP for the required frequency bands are not unique and several sizes may result in similar frequency responses.

lines that pass from the origin with positive slopes. The slopes can be controlled by changing the sizes of patches; larger patches with smaller separations have larger slopes. Addition of $Y_{s2}$ can change $B_{s1,req}^{TE}$ from a convex function of frequency at both frequency bands to a convex function at 2 GHz and a concave function at 8 GHz. Actually, for single layer RMCPs the concave parts of required $B_{s1}^{TE}$ always have negative values [26]. Therefore, the $B_{s1}^{TE}$ of patches cannot pass from them, since the patches always produce positive values of $B_{s1}^{TE}$. Consequently, the second layer FSS is required to make the concave part of required $B_{s1}^{TE}$ positive at the second operating frequency. This will help to construct the dual-band RMCP as will be indicated shortly.

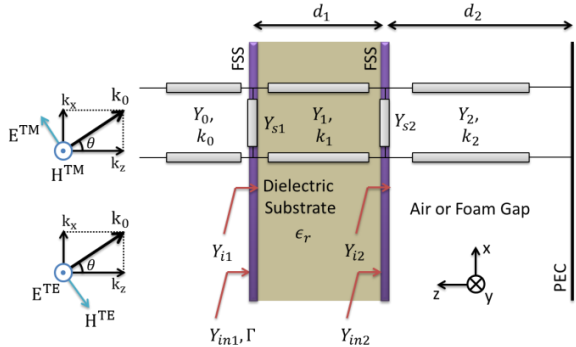

Fig. 1. Equivalent transmission line model for the RMCPs with dual-layer FSS.

RHCP and LHCP reflected waves can be obtained from the incident wave with polarization angle of 135°, if we have $\Gamma_{req}^{TE} = \pm j\Gamma^{TM}$, since the reflected TE and TM waves should have identical amplitudes and $\pm 90°$ phase difference. Electric fields of the incident and reflected waves with TE and TM components are indicated in Fig. 2(c). Electric field of the reflected wave rotates as the time elapses, however it is indicated at one point as an example. Using the equivalent transmission line model in Fig. 1, both TE and TM reflection coefficients are given by $\Gamma = (Y_0 - Y_{in1})/(Y_0 + Y_{in1})$, where $Y_{in1} = Y_{i1} + Y_{s1}$ with the corresponding admittances. Now by replacing TE and TM reflection coefficients in the equation $\Gamma_{req}^{TE} = \pm j\Gamma^{TM}$, the required TE admittance will be [20]

$$Y_{s1,req}^{TE} = jB_{s1,req}^{TE}$$
$$= j\left[\frac{-j(Y_{i1}^{TM} + Y_{s1}^{TM}) \mp Y_0^{TM}}{Y_0^{TM} \mp j(Y_{i1}^{TM} + Y_{s1}^{TM})} Y_0^{TE} + jY_{i1}^{TE}\right], \quad (1)$$

with $Y_i^{TM} = \omega\epsilon_0\epsilon_{ri}/k_{zi}$ and $Y_i^{TE} = k_{zi}/\omega\mu_0$, where $\epsilon_{ri}$ is the relative permittivity of the $i$th layer. Also, $k_{zi} = \sqrt{k_i^2 - k_x^2}$, where $k_i = \sqrt{\epsilon_{ri}}k_0$. Here we have three layers with $\epsilon_{r0} \cong \epsilon_{r2} \cong 1$ and $\epsilon_r = 3.9$. $k_x$ is unique and equals $k_0 \sin\theta$ in all layers due to the boundary conditions. The real part of admittance in neglected with ignoring losses and only the susceptance, $B_{s1}^{TE}$, is assumed. In addition, it should be considered that $Y_{in2} = Y_{i2} + Y_{s2}$, in calculating $Y_{i1}$ for both TE and TM modes [26].

Equations of the susceptances of FSS with rectangular patches are not repeated since they are available in previous reports [26, 28]. However, as functions of frequency they are

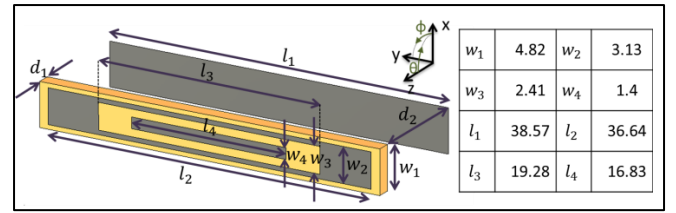

(a)

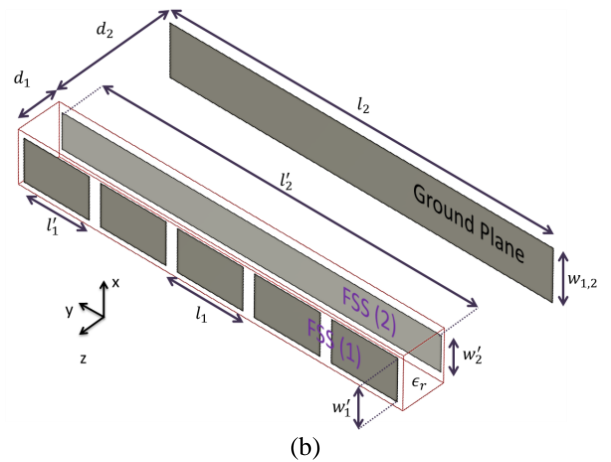

(b)

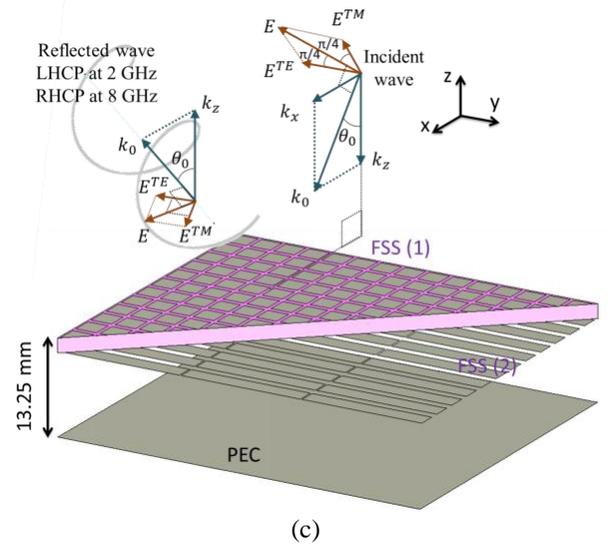

(c)

Fig. 2. Schematic of the dual band RMCPs; (a) unit-cell of the RMCP with single layer FSS [26] (dimensions in mm), (b) unit-cell and (c) array with slashed first layer of the proposed RMCP with dual-layer FSSs.

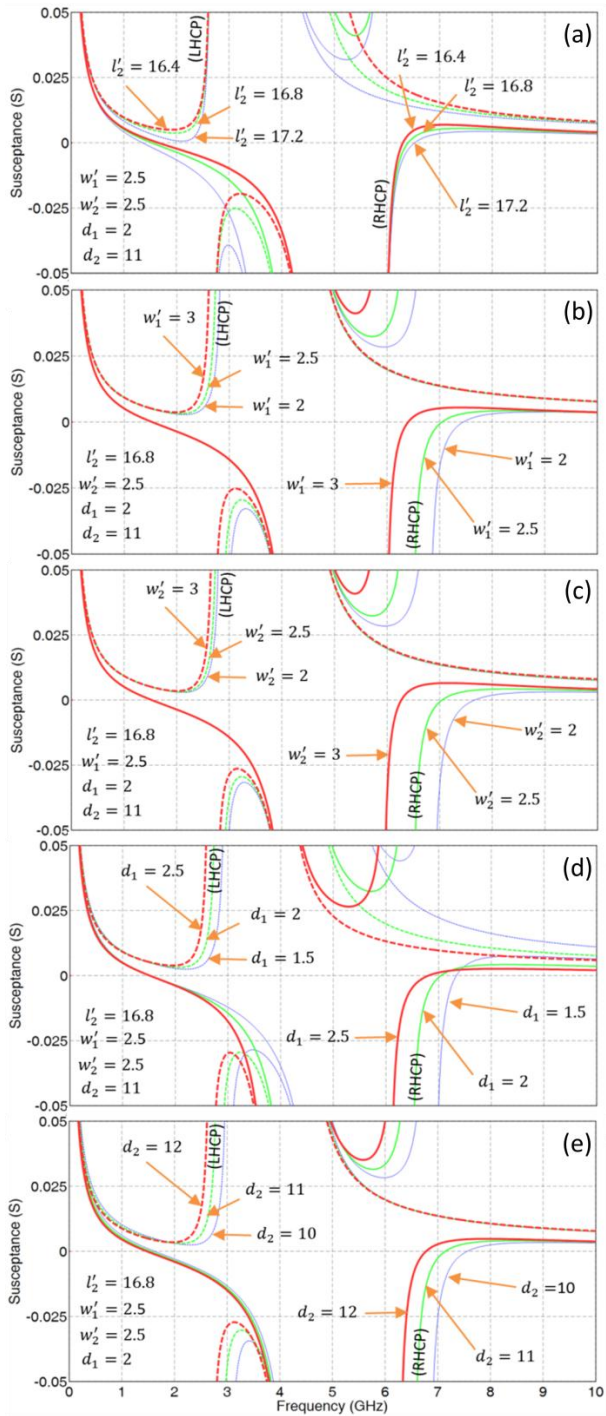

Fig. 3. Parametrical study on the required TE susceptance of the first layer patches for CP; the effects of changing (a) $l'_2$, (b) $w'_1$, (c) $w'_2$, (d) $d_1$ and (e) $d_2$ on the required TE susceptance respectively.

The effects of changing dimensions of the proposed RMCP in Fig. 2(b) on the required susceptance are indicated in Fig. 3. Maximum effects for LHCP at S band are for $l'_2$ while it does not shift the frequencies significantly as indicated in Fig. 3(a). The effects of changing some parameters are similar, such as $w'_1$ and $w'_2$ as can be observed in Fig. 3(b) and Fig. 3(c). They can be used to adjust the separation between operating frequencies since the frequencies are shifted by changing them. Both parameters have more impacts on the X band susceptance; however the effects of changing $w'_1$ are a bit larger on the required $B^{TE}_{s1}$ for LHCP at S band while the effects of changing $w'_2$ are somewhat larger for RHCP at X band. The required susceptance at X band can be adjusted by changing $d_1$ as indicated in Fig. 3(d), most significantly. In addition, it can be observed in Fig 3(e) that both frequencies can be shifted by changing $d_2$.

As a consequence of these diagrams one has multiple choices for the dimensions to produce the desired responses. Initial dimensions of the proposed RMCP based on this parametrical study are given in Fig. 4. In addition, it is indicated that if we have $l'_1 = 3.2$ mm, the calculated $B^{TE}_{s1}$ should meet the required $B^{TE}_{s1}$ at the desired frequencies. Also, there are multiple choices for the sizes of patches. For example, if we have assumed that the unit-cell of second layer is four times greater than the first layer in $\hat{x}$ direction ($l_1 = 4.625$), the $B^{TE}_{s1}$ curve for $l'_1 = 3.975$ mm would be exactly the same.

Effects of changing the propagation angle ($\theta$) on the susceptances are also indicated in Fig. 4. These effects on the phase difference between calculated TE and TM reflection coefficients ($\angle\Gamma^{TE} - \angle\Gamma^{TM}$) are indicated in Fig. 5(a). It can be observed that the increase of $\theta$ should increase the S band bandwidth and reduce the X band bandwidth and vice versa. This can be recognized from Fig. 4 where the required $B^{TE}_{s1}$ is a convex function of frequency at S band and concave function at X band despite the fact that it is a convex function for both frequency bands of the single layer RMCP [26]. Phase differences are of the simulated design are also indicated in Fig. 5(b) which imply similar behavior with calculations as will be discussed in next section.

### III. SIMULATION AND MEASUREMENT RESULTS

Simulated axial ratio (AR) of the reflected wave from the proposed RMCP is indicated in Fig. 6 using frequency domain solver in CST. Simulated phase differences between TE and TM reflection coefficients are also indicated in Fig. 5(b). The initial dimensions are similar with the dimensions in Fig. 4. Simulation results of the RMCP with initial sizes indicate dissimilarities with the predictions in Fig. 4 and Fig. 5(a). It can be interpreted as the coupling effects between layers and the higher order expressions in calculating the admittances of FSS layers [28], which are neglected in the transmission line model. It appears that $B^{TE}_{s1}$ does not cross the required $B^{TE}_{s1}$ at X band. This can be compensated by reducing $l'_1$ to decrease the slope of TE susceptance of first layer patches, as can be observed in Fig. 4. Therefore AR of the RMCP with reduced $l'_1$ from 3.2 mm to 3 mm is indicated in Fig. 6. It can be observed reducing $l'_1$ affected the S band AR as well. However, this can be compensated by increasing $l'_2$ as can be observed in Fig. 3. Consequently, one can obtain desired results by slightly changing the dimensions using the understanding of each parameter's effect on the required susceptances. We changed only two parameters, $l'_2$ from 17.2 mm to 17.25 mm and $l'_1$ from 3.2 mm to 3 mm. Simulated AR of the modified RMCP in Fig. 6 indicates 44 % bandwidth for S band and 46 % bandwidth for X band for the incident wave with $\theta = 45°$.

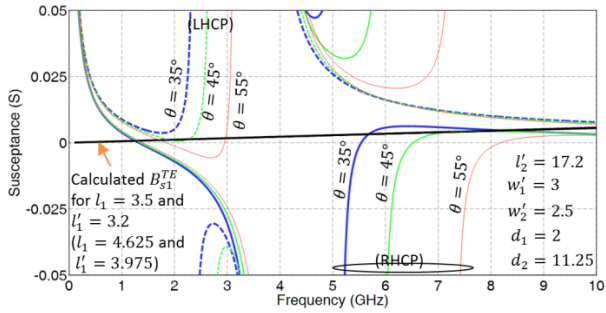

Fig. 4. Required $B_{s1}^{TE}$ for CP with determined dimensions ($l'_2$, $w'_1$, $w'_2$, $d_1$ and $d_2$) and the calculated $B_{s1}^{TE}$ for two equivalent dimensions of $l_1$ and $l'_1$.

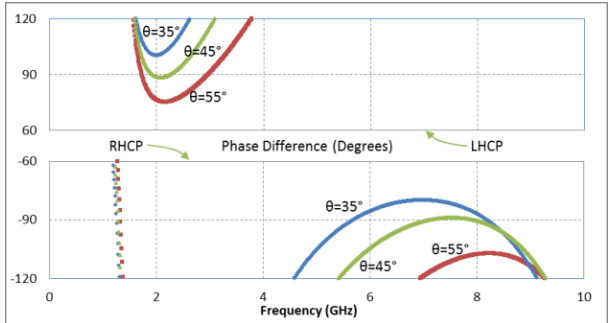

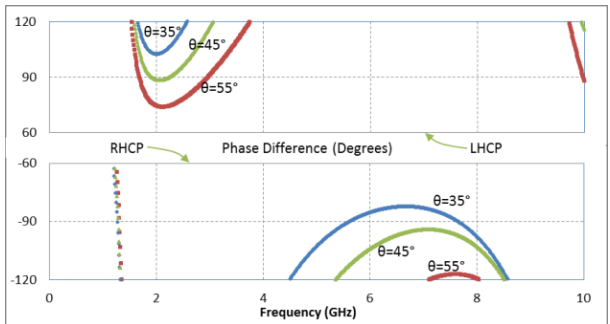

Fig. 5. Phase differences between TE and TM reflection coefficients of the RMCP with initial dimensions for different propagation angles; (a) calculated phase differences and (b) simulated phase differences.

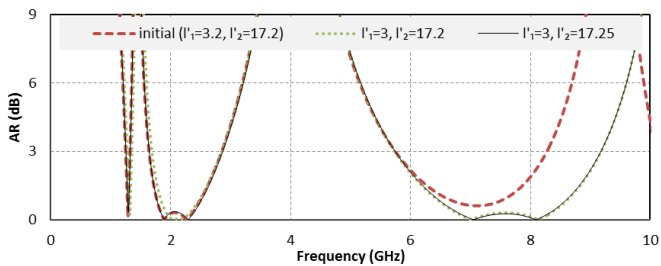

Fig. 6. Simulated AR of the reflected wave from the RMCP with the initial sizes from calculations and modified sizes, the incident wave with 135° polarization angle.

Simulated AR of the reflected wave from proposed RMCP is indicated in Fig. 7 for different angles of the incident wave. Comparison between ARs of the proposed RMCP and the single-layer RMCP [26] is indicated in Fig. 8 for two angles of the incident wave as the examples. It can be observed that the permitted incident angle difference for X band of this RMCP is improved in comparison with the single layer RMCP [26]. In addition, there is a tradeoff between the bandwidths of X band and S band that can be controlled by maximum and minimum propagation angles of the incident wave. For example if 15° angle difference is required, the bandwidth will be 30.5 % and 33.7 % for S band and X band respectively with $\theta_{max} = 50°$ and $\theta_{min} = 35°$. If we increase $\theta_{max}$ to 55° the X band bandwidth will reduce to 12.2 %. Same angle difference can be obtained by reducing $\theta_{min}$ to 30° that will reduce the S band bandwidth to 14.2 %. Therefore we can increase the angle difference by reducing the bandwidth of only one operating frequency band. Furthermore, the permitted angle difference can be improved by reducing the bandwidths. For example, the angle difference can be improved up to 25° with the bandwidths 14.2 % and 12.2 % for S and X bands respectively.

Simulations are validated using measured results of a 297.5 mm × 341.3 mm prototype. The realized RMCP and the measurement setup are indicated in Fig. 9. The 11.25 mm separation between the aluminum plate and the substrate is provided using pieces of foams with negligible permittivity. Transmitter antenna is placed with 135° polarization angle and the polarization of receiver is change in multiple angles as indicated in Fig. 9. AR is the ratio of highest received power and lowest power between the received powers in all polarization angles. Measured AR of the reflected wave from the RMCP with three angles of incident wave ($\theta$ in Fig. 9) is indicated in Fig. 10. It can be observed that the measured results are in agreement with the simulations. Finally, simulated and measured reflection coefficients of the proposed RMCP are indicated in Fig. 11 at $\theta = 45°$. Small fluctuations are predictable in measurement due to the low level noise around the receiver. However, it can be observed that the reflection loss is less than 0.4 dB throughout the bandwidths.

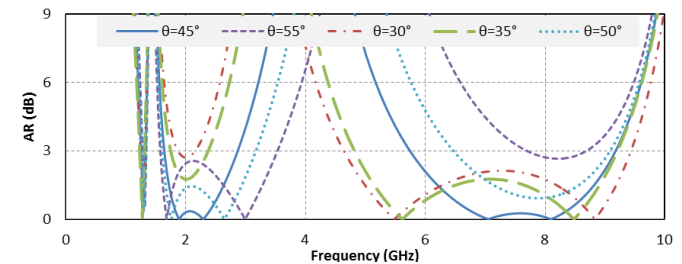

Fig. 7. Simulated AR of the reflected wave with incident wave with 135° polarization and different propagation angles.

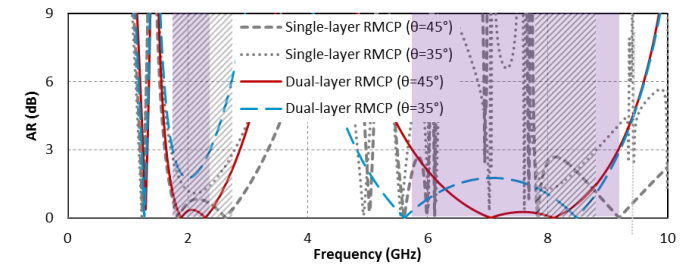

Fig. 8. Comparison between ARs of the proposed RMCP and the single-layer RMCP [26]; operating frequencies of the single layer RMCP is indicated by diagonal hatches and the dual-layer RMCP by solid fills.

## IV. CONCLUSION

A dual band reflection-mode linear to circular polarization converter is proposed in this paper. The proposed RMCP consists of dual-layer rectangular patches. Required TE susceptance of the first layer is calculated in multiple situations and a parametrical study is applied on the susceptance. Dimensions of the patches are tuned for the best operation on 2 GHz and 8 GHz. The AR bandwidth is 30.5 % and 33.7 % for first and second frequency bands respectively with $\theta_{max} = 50°$ and $\theta_{min} = 35°$. The AR bandwidth can be improved by reducing $\Delta\theta$. In addition, bandwidth of the first operating frequency can be improved by reducing the bandwidth of second operating frequency and vice versa. Furthermore, the permitted angle difference can be improved by reducing the bandwidths. Finally, a 297.5 mm × 341.3 mm prototype of the proposed dual band RMCP is fabricated and tested successfully. Measurement results indicate good agreements with simulations at different angles of the incident wave. In addition, less than 0.4 dB reflection loss is observed in measurement at both frequency bands.

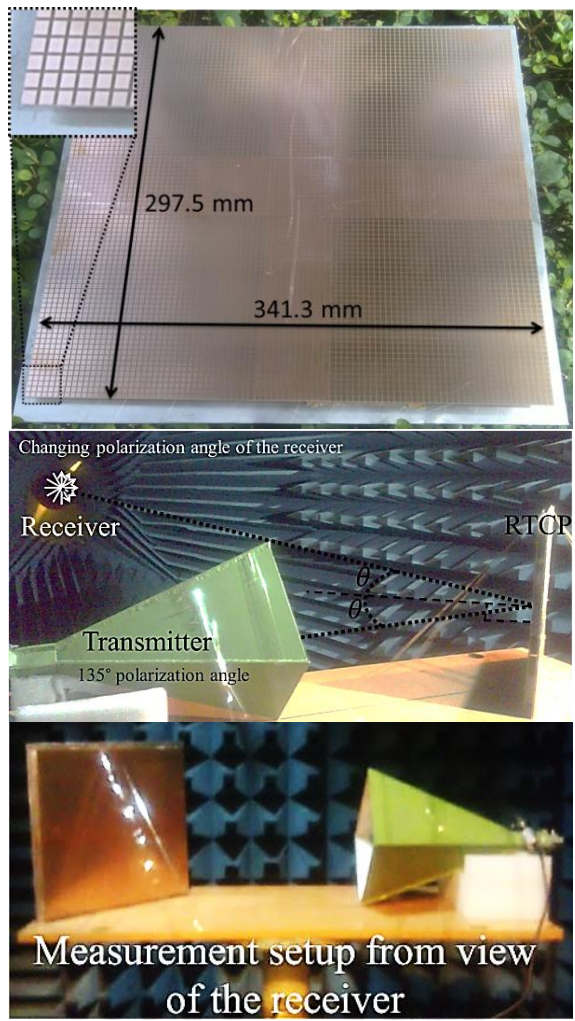

Fig. 9. Photograph of the fabricated RMCP and measurement setup from two viewangles.

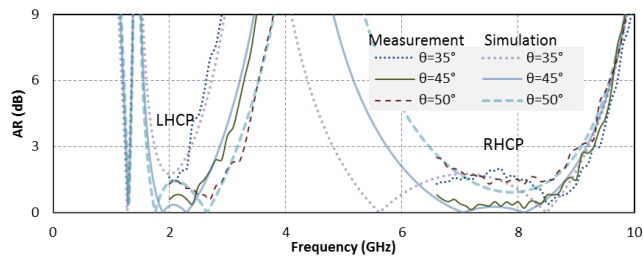

Fig. 10. Measured AR of the reflected wave from incident wave with 135° polarization at different propagation angles compared with simulation results.

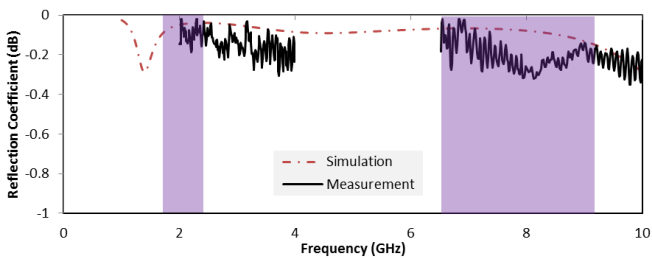

Fig. 11. Simulated and measured reflection coefficients of the proposed RMCP at $\theta = 45°$.